\documentstyle[11pt,newpasp,twoside,epsf,psfig]{article}
\def\b#1{{\bf #1}}
\def\pc{\,{\rm pc}}\def\kpc{\,{\rm kpc}}
\def\Gyr{\,{\rm Gyr}}\def\pa{\partial}
\def\fracj#1#2{{\textstyle{#1\over#2}}}
\def\kms{\,{\rm km}\,{\rm s^{-1}}}
\begin{document}
\title{Secular Evolution of the Galactic Disk}
 \author{James Binney}
\affil{Oxford University, Theoretical Physics, Keble Road, Oxford, OX1 3NP,
U.K.}

\begin{abstract}
In the solar-neighbourhood, older stars have larger random velocities than
younger ones. It is argued that the increase in velocity dispersion with
time is predominantly a gradual process rather than one induced by discrete
events such as minor mergers. Ephemeral spiral arms seem to be the
fundamental drivers of disk heating, although scattering by giant molecular
clouds plays an important moderating role. In addition to heating the disk,
spiral arms cause stars' guiding centres to diffuse radially. The speed of
this diffusion is currently controversial.

Data from the Hipparcos satellite has made it clear that the Galaxy is by no
means in a steady state. This development enormously increases the
complexity of the models required to account for the data. There are
preliminary indications that we see in the local phase-space distribution
the dynamical footprints of long-dissolved spiral waves. 
\end{abstract}

\section{Introduction}

It is now half a century since Roman (1950) and Parenago (1950) discovered
that the kinematic properties of stars near the Sun are strongly correlated
with spectral type. The theory of stellar evolution soon showed that the
sense of this correlation was that older stellar groups have larger velocity
dispersions and asymmetric drifts than younger groups. There are two generic
explanations for this phenomenon. In one picture the turbulent velocities in
the gas from which stars form has declined steadily since the Galaxy started
to form, and the current random velocities of stars are fossil records of
the turbulent velocities at the moment of their formation. This picture,
which inspired the classic paper of Eggen Lynden-Bell \& Sandage (1962), is
now not widely favoured, although it still has proponents (e.g., Burkert,
Truran \& Hensler, 1992).  In the other picture the random velocities of
stars are small ($\sim7\kms$) at birth and increase with time. Whether this
increase is continuous or episodic in nature is currently being debated.

\begin{figure}
\centerline{\psfig{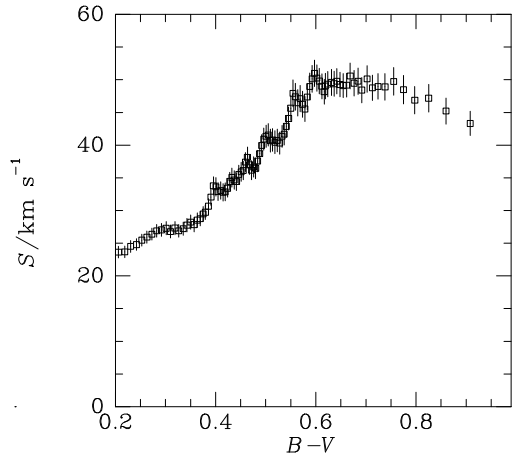}}
\caption{Random velocity on the sky versus colour for main-sequence stars
with good Hipparcos parallaxes. [From Binney, Dehnen \& Bertelli (2000)]\label{DB1}}
\end{figure}

\section{The Age Velocity-Dispersion Relation}

Fig.~\ref{DB1} is a plot of the random velocity on the sky, $S$, of a
kinematically unbiased sample of stars in the Hipparcos Catalogue that have
good parallaxes. As one proceeds from blue to red stars, $S$ rises steadily
until one reaches $(B-V)=0.6$, where it abruptly levels off. For
$B-V\ga0.75$, $S$ gently declines. The abrupt change in slope at
$(B-V)=0.6$ is called Parenago's discontinuity, and the natural
interpretation of this phenomenon is this. Bluewards of the discontinuity
stars have main-sequence lifetimes shorter than the age of the solar
neighbourhood, $\tau_{\rm max}$, while redwards of it lifetimes exceed
$\tau_{\rm max}$. Consequently, any tendency of the velocity dispersion of a
stellar group to increase over time will cause $S$ to increase with $B-V$ at
$(B-V)<0.6$ because in this range the age of the oldest stars
contributing to $S$ increases with $B-V$. Conversely, $S$ should be
independent of $B-V$ redward of the discontinuity.

Binney, Dehnen \& Bertelli (2000; BDB) show that the structure of
Fig.~\ref{DB1} can be accurately reproduced if the velocity dispersion of a
stellar group increases with age as $\tau^{0.33}$. Interestingly, their
models reproduce the decline in $S$ at the reddest colours, which is not
predicted by the simple considerations of the last paragraph.  The origin of
this decline is a change in the age distribution with colour that arises
because near $(B-V)=0.6$ the oldest stars are turning off the ZAMS, with
the result that a magnitude-limited sample of stars is biased towards these
stars. At significantly redder $B-V$, even the oldest stars are still close
to the ZAMS, and the sample contains a fair sample of all the stars ever
formed. 

The BDB fits to Fig.~\ref{DB1} formally require the age of the solar
neighbourhood to be $11.2\pm0.75\Gyr$, which is older than current
cosmological fashions lead one to expect. In fact, BDB show that the
structure of Fig.~\ref{DB1} can be even better reproduced if one adopts an
age $\tau_{\rm max}=9\Gyr$. What makes such young ages unacceptable is the
distribution of stars within the sample over colour: models with $\tau_{\rm
max}=9\Gyr$ are deficient in red stars. Since the selection function of the
sample is not precisely defined, this objection may not be serious. A more
serious objection may be that young ages require rather flat slopes $\alpha$
of the IMF: whereas $\tau_{\rm max}=11.2\Gyr$ requires a slope
$\alpha=2.25\pm0.5$ that is indistinguishable from Salpeter's slope (2.35),
$\tau_{\rm max}=9\Gyr$ requires $\alpha=1.75\pm0.5$.

\begin{figure}
\centerline{\psfig{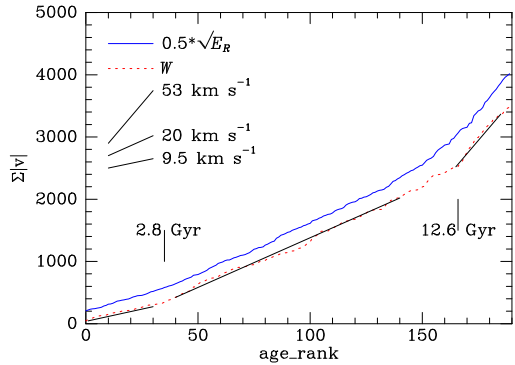}}
\caption{Cumulative velocity versus age rank for the 189 F stars in the
Edvardsson et al.\ (1993) sample. The lower curve shows the partial
$\sum|W|$ of speeds perpendicular to the plane, while the upper curve shows
the partial sum of square roots of the stars' epicycle invariants
$\fracj12(U^2+\gamma^2V^2)$, where $\gamma=2\Omega/\kappa$. [After Freeman (1991)]
\label{KF}}
\end{figure}
 
Fig.~\ref{DB1} is based on over $12\,000$ stars but its interpretation
requires considerable modelling. We can obtain less model-dependent
information about the age-velocity dispersion relation by examining much
smaller samples of stars for which individual age estimates are available.
Fig.~\ref{KF} is for the data set that is currently the best available,
namely the 189 F stars which Edvardsson et al.~(1993) observed in great
detail. The stars have been ordered by age and the vertical axis shows the
sum of the speeds of the stars that are younger than the star of a given
rank, which is plotted horizontally. If the speeds of stars were drawn from
a single Gaussian of dispersion $\sigma$, regardless of rank, $\sum|W|$
would on average increase with rank $n$ as $(2/\pi)^{1/2}\sigma n$. Freeman
(1991) pointed out that the $W$ curve in Fig.~\ref{KF} appears to have three
linear sections, with slopes that correspond to $\sigma=9.5$, $20$ and
$53\kms$.\footnote{Actually, Freeman gave the highest velocity dispersion as
$42\kms$, but this appears to be an error.} Thus, the suggestion is that the
vertical velocity dispersion of stars increases from $9.5$ to $20\kms$ on a
timescale $\sim3\Gyr$, and is then constant at $20\kms$. The oldest stars
($\tau\ga12.6\Gyr$) form the thick disk and have significantly larger
velocity dispersions. The abrupt increase in velocity dispersion from $20$
to $53\kms$ could well have occurred when a low-mass satellite swept through
the disk in the manner described by Quinn \& Goodman (1986).  Quillen
\& Garnett (2000) have reaffirmed these conclusions after reanalyzing the
Edvardsson et al.\ sample using Hipparcos parallaxes and up to date
stellar-evolution models. The up-to-date stellar ages bring down to
$\tau\sim10\Gyr$ the age at which the velocity dispersion rises to $53\kms$.

This picture is an intriguing one that merits careful consideration, but one
does have to be cautious because the sample is small and not unbiased.
Moreover, when one attempts to model the data of Fig.~\ref{DB1} under the
assumption that the velocity dispersion increases smoothly up to an age of
$8$ or $10\Gyr$ and then jumps to a constant final value, one finds that the
best fitting model has a negligible discontinuity in velocity dispersion.
For example, if the jump is assumed to occur at $t=8\Gyr$, $S$ {\em drops\/}
from $55$ to $52\kms$ at that time.  The large Hipparcos sample simply does
not support the implication of the Edvardsson sample that the velocity
dispersion changes abruptly. I shall henceforth assume that velocity
dispersion increases continuously over time.

\section{Heating mechanisms}

Stellar disks are fragile objects because their distribution functions crowd
all stars into a low-dimensional structure in phase space. Any perturbation
is liable to increase the disk's entropy by  scattering stars out into the
body of phase space. It is worth noting that the perturber does not need to
supply energy; all it has to do is to scatter stars onto more eccentric
orbits and/or orbits that are inclined to the disk's equatorial plane.

Spitzer \& Schwarzschild (1953) suggested that stars were scattered by gas
clouds. This was a visionary proposal because clouds of sufficient mass and
compactness would not be discovered for 20 years. When the Galaxy was
studied in the $2.6\,$mm line of CO, the existence of giant
molecular clouds (GMCs) with the predicted masses was established, and it
was widely assumed that GMCs were responsible for heating the disk. Lacey
(1984) cast doubt on this conclusion for two reasons. First, he showed that
GMCs rapidly established a characteristic ratio
$\sigma_z/\sigma_R\simeq0.78$ between the vertical and radial velocity
dispersions, whereas the observed ratio is significantly smaller: $0.6$
(Dehnen \& Binney, 1998). Second, Lacey showed that the efficiency of
heating by GMCs declines in time more rapidly than had been thought, with
the result that GMCs cannot accelerate stars to the largest observed
dispersions. 

The slowing of acceleration by GMCs with increasing velocity dispersion is
easy to understand. There are two aspects. First the cross-section to
Coulomb scattering through a given angle of stars by Clouds falls off with
encounter speed $v$ as $\sim v^{-2}$. Second, the GMCs are confined to a
thin sheet, in which stars spend less and less time as their random
velocities increase.  In the idealized two-dimensional scattering problem
considered by Spitzer \& Schwarzschild (1953), clouds increase the stellar
velocity dispersion as $\sigma\sim t^\beta$, with $\beta=1/3$.  When motions
perpendicular to the plane are included, one finds $\beta=1/4$ in the case
that the vertical oscillations of stars are harmonic. In the more realistic
case of anharmonic vertical oscillations, a still smaller value of $\beta$
is appropriate (Jenkins, 1992).

Sellwood \& Carlberg (1984) revived the proposal of Barbanis \& Woltjer
(1967) that spiral arms heat the disk. Lynden-Bell \& Kalnajs (1972) had
shown that a wave heats only those stars that resonate with it. Hence to
heat the entire disk one needs either a wave whose frequency sweeps over a
wide range, or many transient waves. Numerical simulations by Sellwood and
others (Sellwood \& Carlberg 1984; Sellwood \&
Kahn, 1991) show that stellar disks are indeed rife with such features.

Spiral arms excite random motions parallel to the plane, but not vertical
oscillations. Hence, one is led to a composite picture in which spiral arms
increase the in-plane dispersions, and GMCs divert the in-plane motion
into vertical oscillations. Binney \& Lacey (1988) established a formalism
within which the two mechanisms could be combined. They focused on the
distribution of stars in the integral space whose coordinates are epicycle
energy, $E_R$, and energy of vertical oscillation, $E_z$. These quantities
are intimately connected with the radial and vertical actions $J_R$ and
$J_z$ -- in the epicycle approximation we have $J_R=E_R/\kappa$ and
$J_z=E_z/\nu_z$, where $\kappa$ is the usual epicycle frequency and $\nu_z$
is the angular frequency of vertical oscillations. If we assume that
the drift of stars through integral space is driven by large numbers of
small and statistically independent disturbances, the evolution of the
phase-space density of stars, $f(\b J)$, can be found by solving the Fokker-Planck
equation
 \begin{equation}
{\pa f\over\pa t}=\fracj12\sum_{ij}{\pa\over\pa J_i}
\left(\Delta_{ij}{\pa f\over\pa J_j}\right),
\end{equation}
 where the diffusion tensor, $\Delta_{ij}=\langle\delta J_i\delta J_j\rangle$, is
the expectation of the product of the changes in $J_i$ and $J_j$ per unit
time and can be evaluated if one knows the statistical characteristics of
the disk's gravitational potential.

As Lacey (1984) showed, clouds have a strong tendency to establish an
equilibrium between oscillations parallel to the plane and vertical
oscillations, while waves only excite oscillations parallel to the plane.
Consequently, the relative effectiveness of clouds and waves can be
gauged from the degree to which the observed ratio $\sigma_z/\sigma_R$
deviates from the clouds' equilibrium value. Jenkins \& Binney (1990) used
this idea to estimate the relative magnitudes of the contributions of clouds
and waves to the diffusion tensor. They found that waves are strongly
dominant: 
 \begin{equation}\label{JBeq}
\Delta_{RR}^{\rm wave}\simeq90\Delta_{RR}^{\rm cloud}.
\end{equation}

\begin{figure}
\centerline{\psfig{file=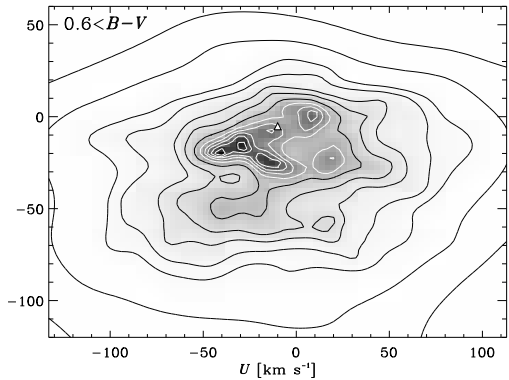,width=.5\hsize}}
\caption{The density of red solar-neighbourhood stars projected onto the
$(U,V)$ plane of velocity space. The small triangle marks the Local Standard
of Rest. [After Dehnen (1998)]\label{Dehnf4}}
\end{figure}

While the jury is still out on this question, the preliminary indications
are that this conclusion is confirmed in that the dynamical footprints of
spiral arms may be visible in the velocity-space distribution of stars near
the Sun that has been deduced from Hipparcos data. Fig.~\ref{Dehnf4} shows,
from the work of Dehnen (1998), the projection onto the $(U,V)$ plane of the
velocity-space density of stars that are redder than Parenago's
discontinuity. Even though the stars in question are mostly more than
$4\Gyr$ old, many local density maxima are apparent and the outer contours
are far from elliptical. Dehnen demonstrates that these features are real
rather than reflections of statistical uncertainty.

If $f$ were a function of actions only, as the distribution function of a
steady-state galaxy should be (Jeans' theorem), the density of stars in
Fig.~\ref{Dehnf4} would be constant on the ellipses
$\hbox{constant}=U^2+\gamma^2V^2$ of constant $J_R$.\footnote{The stars upon
which Fig.~\ref{Dehnf4} is based lie up to $\sim100\pc$ from the Sun, and in
the formula for $J_R$ we ought strictly to use not $V$ but the velocity of
the star relative to its own LSR, which differs from that of the Sun by
$-2A(R-R_0)$, where $A$ is Oort's constant. This correction typically
amounts to less than $3\kms$, however.} From the fact that the stellar
density in Fig.~\ref{Dehnf4} varies markedly around these ellipses, it is
clear that $f$ depends on angle variables as well as actions, and that the
Galaxy is not in a steady state.

In the bottom left-hand corner of Fig.~\ref{Dehnf4} there is a long
peninsula of high density that reaches out to $(U,V)=(-45,-50)\kms$. Raboud
et al.\ (1998) and Dehnen (2000) argue that this feature is generated by the
Galactic bar, which places stars on highly eccentric orbits that pass the
Sun as they move outwards in their approach to apocentre. Some of the
fine-scale structure closer to the LSR (marked by a small triangle) is
associated with star clusters such as the Hyades, but it seems unlikely that
all structure can be explained in this way. De Simone \& Tremaine (2000)
suggest that much of this structure is generated by spiral arms. They
calculated the final star density in the $(U,V)$ plane when a series of
externally imposed and uncorrelated spiral waves acted on an initially cool
stellar disk. Their densities show local density maxima just as
Fig.~\ref{Dehnf4} does, but they have more maxima than the observational
plot does. It is unclear whether this excess reflects inadequate
resolution in the observational data, or  weaknesses in the simulations'
spiral arms.

\section{Radial migration}

When a star is scattered, whether by spiral arms or by molecular clouds, it
is liable to change its angular momentum, $L$, about the Galaxy's vertical
axis, as well as its radial and vertical actions. The guiding centre of a
star's orbit is just $L/v_c$, so changes in $L$ are directly associated with
radial migration. Radial migration can in principle be detected because
there is a metallicity gradient within the disk (Wielen, Fuchs \& Dettbarn
1996; WFD). In fact, it is widely believed that  all interstellar
material at a given time and radius has a common metallicity $Z(R,t)$
(Edmunds, 1998). By contrast, we know from the work of Edvardsson et al.\
(1993) that there is considerable scatter in the metallicities of stars that
have a common guiding centre and age. Does this scatter arise because these
stars were born at different radii, where interstellar gas had different
metallicities at their common time of formation? 

In this context the Sun, which is currently near pericentre, so its guiding
centre lies $\sim200\pc$ outside $R_0$, provides an interesting case study.
It is more metal-rich than the local average for stars of its age by
$0.17\,$dex, and more metal-rich than gas in the local Orion star-forming
region by $0.47\,$dex.  These observations are remarkable, given that the
mean metallicity of both the stellar and gaseous disks decrease outwards and
one usually imagines that the metallicity of the ISM tends to increase over
time,\footnote{In the presence of accretion, there is no guarantee
that $Z$ will increase with time at a given radius.} so the local ISM should
have been even more metal-poor than Orion $4.5\Gyr$ ago, when the Sun
formed.  WFD ask, at what radius did the ISM have metallicity $Z_\odot$ when
the Sun formed, and could the Sun's guiding centre have migrated from that
radius to its present location as a result of a series of uncorrelated
scattering events?  They conclude that the Sun formed at
$R_0-2\kpc\simeq6\kpc$ and its guiding centre has since migrated outwards to
$R_0+200\pc$.

De Simone \& Tremaine (2000) have calculated the extent of radial migration
in their simulations of star-wave scattering and obtain an answer,
$700\pc\la\Delta R\la1200\pc$, that is smaller than that 
derived by WFD. Unfortunately, there is no guarantee that the spiral
features simulated by De Simone \& Tremaine were sufficiently realistic.

\section{Conclusions}

Jeans' theorem simplifies galactic dynamics greatly because it requires the
distribution function of a steady-state galaxy to be a function of only
three rather than six variables. This simplicity is so alluring that one
tries to imagine the Galaxy as slowly moving from one steady state to
another. An orbit-averaged Fokker-Planck equation describes this evolution,
and has proved reasonably successful in accounting for the dependence upon
age of measures of the kinematics of stars such as velocity dispersions and
asymmetric drifts. Moreover, an orbit-averaged Fokker-Planck equation
provides a unified framework in which to discuss the effects of ephemeral
spiral arms and GMCs, which act together to drive the observed evolution of
the solar neighbourhood.

Ephemeral spiral arms steadily drive upwards the radial and tangential
components of the velocity-dispersion tensor, while GMCs deflect stars out
of the plane, so that $\sigma_z$ increases roughly in step with $\sigma_R$,
although not so fast that equilibrium is established between these two
dispersions. The guiding centres of stars gradually diffuse in radius,
although the distance typically travelled is currently controversial:
estimates range from $\sim700\pc$ up to $\sim2\kpc$.

When the phase-space distribution of stars is examined in detail one finds
that there is a great deal more structure than can be described within the
context of Jeans' theorem and an orbit-averaged Fokker-Planck equation.
Evidently, the Galaxy is significantly displaced from a steady state.

At one level this discovery is a disappointment, since it enormously
increases the complexity of the models required to account for the data. At
another level it is a tremendous opportunity to learn more about the
structure and the history of the Galaxy. We should seize this opportunity
with both hands because we will probably never have comparably rich data for
any other Galaxy. As yet there is no complete interpretation of the
fine-scale structure seen in the local phase-space density, but there is a
strong case that some of it reflects the existence of the Galactic bar, and
a weaker case that other structure is generated by the spiral waves that
dominate the heating of the disk.

\section*{Acknowledgments}

I thank S.D.~Tremaine for communicating work prior to
publication and the University of Washington for its hospitality during the
writing of this article. This work was supported in part by NSF grant
AST-9979891.

\end{document}